\documentclass[aps,superscriptaddress,prl,nofootinbib,reprint]{revtex4-1}

\usepackage[ colorlinks = true,
             linkcolor = blue,
             urlcolor  = blue,
             citecolor = blue,
             anchorcolor = green,
]{hyperref}
\usepackage{subfigure}
\usepackage{color}
\usepackage{tabularx}
\usepackage{amsfonts}
\usepackage{amssymb}
\usepackage{amsthm}
\usepackage{amsmath}
\usepackage{amsxtra}
\usepackage{graphics}
\usepackage{enumerate}
\usepackage{tikz}

\def\ket #1{\vert #1\rangle}

\def\tr{{\rm Tr}}

\newcommand*{\cB}{\mathcal{B}}

\newcommand*{\cH}{\mathcal{H}}
\newcommand*{\cI}{\mathcal{I}}
\newcommand*{\cJ}{\mathcal{J}}
\newcommand*{\cK}{\mathcal{K}}

\newcommand*{\cM}{\mathcal{M}}

\newcommand*{\cD}{\mathcal{D}}

\newcommand*{\cR}{\mathcal{R}}

\newcommand*{\cT}{\mathcal{T}}
\newcommand*{\cV}{\mathcal{V}}
\newcommand*{\cW}{\mathcal{W}}

\newcommand{\bc}{\begin{center}}
\newcommand{\ec}{\end{center}}

\def\01{\{0,1\}}

\usepackage{mathtools}
\mathtoolsset{showonlyrefs}

\newtheorem{lemma}{Lemma}

\newtheorem{theorem}{Theorem}
\newtheorem*{theorem*}{Theorem}

\begin{document}

\title{Continuum limits of homogeneous binary trees and the Thompson group}

\author{Alexander Kliesch}
\affiliation{Zentrum Mathematik, Technische Universit\"at M\"unchen, 85748 Garching, Germany}
\email{kliesch@ma.tum.de}

\author{Robert K\"onig}
\affiliation{Institute for Advanced Study \& Zentrum Mathematik, Technische Universit\"at M\"unchen, 85748 Garching, Germany}
\email{robert.koenig@tum.de}

\newcommand*{\diff}{\mathsf{diff}}
\newcommand*{\thompson}{\mathsf{T}}

\begin{abstract}
Tree tensor network descriptions of critical quantum spin chains are  empirically known to reproduce correlation functions matching CFT predictions in the continuum limit. It is natural to seek a more complete correspondence, additionally incorporating dynamics. On the CFT side, this is determined by a representation of  the diffeomorphism group of the circle. In a remarkable series of papers, Jones outlined a research program where the Thompson group~$\thompson$ takes the role of the latter in the discrete setting, and representations of~$\thompson$ are constructed from certain elements of a subfactor planar algebra. He also showed that for a particular example of such a construction, this approach only yields -- in the continuum limit --  a representation  which is highly discontinuous and hence  unphysical. Here we show that the same issue arises generically when considering tree tensor networks: the set of coarse-graining maps
yielding discontinuous representations  has full measure in the set of all isometries. This extends Jones' no-go example to typical elements of the so-called tensor planar algebra. We also identify an easily verified necessary condition for a continuous limit to exist. This singles out a particular class of tree tensor networks. Our considerations  apply to recent approaches for introducing dynamics in holographic codes.
\end{abstract}

\maketitle

Tensor network techniques  are remarkably successful at capturing the essential features of correlations in many-body quantum systems. The most celebrated examples include Matrix Product States (MPS)~\cite{fannesnachtergaelewerner1992,Wolf:2007wt,Hastings:2007vw},  Projected Entangled Pair States~\cite{schuch2011classifying} or the Multi-scale Entanglement Renormalization Ansatz (MERA)~\cite{PhysRevLett.99.220405}: these variational methods have been successfully applied to (bosonic) spin- and fermionic systems. Variants include tensor networks for quantum field theories~\cite{2010PhRvL.105y1601H,2013PhRvL.110j0402H,jennings2015continuum,2010PhRvL.105z0401O,2010PhRvL.104s0405V}, quantum Hall states and conformal field theories~\cite{nielsenetal12,bernevigetal,koenigscholzprl,zaletelmong}, 
 as well as for non-abelian anyons~\cite{koenigbilgin,pfeiferetal}.

Beyond their descriptive power, tensor networks may help to elucidate the nature  of continuum limits of spin (or anyonic) chains. While the correspondence between critical quantum systems and associated conformal field theories is well-understood in physical terms  -- especially scale-invariance at criticality -- a rigorous construction of CFTs based on a continuum limit of a discrete spin system does not exist, although there are promising first steps in this direction: for example, in certain critical anyonic chains, a subset of observables were shown to converge to operators satisfying the generator relations of the Virasoro algebra in~\cite{ziniwang}.  Tensor networks may contribute to this problem by providing a versatile tool for describing spin chains. Indeed, this is the motivation of our work.

Tree tensor networks give compact descriptions of many-body states with algebraically decaying correlations, and may thus be used to extract critical exponents using efficient optimization algorithms~\cite{silvietal10}. Similar in structure to scale-invariant MERA states, they have a much simpler description and constitute a paradigmatic example of a real-space renormalization process. The basic structure of such a tensor network is illustrated in Fig.~\ref{fig:homogeneousbinarytree}. The correlation functions are determined by an isometry or ``spin-doubling map''~$V:\mathbb{C}^d\rightarrow\mathbb{C}^d\otimes\mathbb{C}^d$. This map gives rise to a coarse-graining procedure and determines critical exponents via derived coarse-graining maps. In particular, an isometry determines an isometric embedding or ``fine-graining map'' $\iota_n:(\mathbb{C}^d)^{\otimes 2^{n}}\rightarrow (\mathbb{C}^d)^{\otimes 2^{n+1}}$ of $2^n$ spins into $2^{n+1}$~spins by $\iota_n=V^{\otimes {2^n}}$. Here $\cH_n:=(\mathbb{C}^d)^{\otimes 2^n}$ may be considered as the Hilbert space associated with degrees of freedom at a length scale defined by $n$. By composing these maps, this also gives  maps $\iota^m_n:\cH_n\rightarrow\cH_m$, embedding $\cH_n$ into $\cH_m$ for $n\leq m$.

\definecolor{darkgreen}{rgb}{0.0, 0.5, 0.0}
\definecolor{bleudefrance}{rgb}{0.19, 0.55, 0.91}
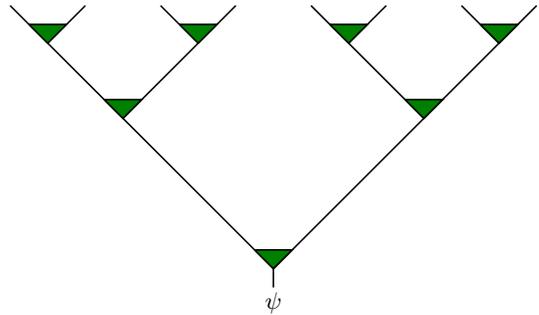
\begin{figure}
\begin{tikzpicture}[scale=0.5]
\node [below] at (0, -0.4) {$\psi$};

\draw [semithick] (-7, 7) -- (0, 0) -- (7, 7); 

\draw [semithick] (-4, 4) -- (-1, 7); 
\draw [semithick] (4, 4) -- (1, 7); 

\draw [semithick] (-6, 6) -- (-5, 7); 
\draw [semithick] (-2, 6) -- (-3, 7); 
\draw [semithick] (2, 6) -- (3, 7); 
\draw [semithick] (6, 6) -- (5, 7); 

\draw [fill=darkgreen, semithick] (-0.5, 0.5) -- (0, 0) -- (0.5, 0.5) -- (-0.5, 0.5);
\draw [semithick] (0, 0) -- (0, -0.5); 

\draw [fill=darkgreen, semithick] (-4.5, 4.5) -- (-4, 4) -- (-3.5, 4.5) -- (-4.5, 4.5);
\draw [fill=darkgreen, semithick] (3.5, 4.5) -- (4, 4) -- (4.5, 4.5) -- (3.5, 4.5);

\draw [fill=darkgreen, semithick] (-6.5, 6.5) -- (-6, 6) -- (-5.5, 6.5) -- (-6.5, 6.5);
\draw [fill=darkgreen, semithick] (-2.5, 6.5) -- (-2, 6) -- (-1.5, 6.5) -- (-2.5, 6.5);
\draw [fill=darkgreen, semithick] (1.5, 6.5) -- (2, 6) -- (2.5, 6.5) -- (1.5, 6.5);
\draw [fill=darkgreen, semithick] (5.5, 6.5) -- (6, 6) -- (6.5, 6.5) -- (5.5, 6.5);
\end{tikzpicture}
\caption{In addition to the isometry~$V:\mathbb{C}^d\rightarrow\mathbb{C}^d\otimes\mathbb{C}^d$ (green triangles), a tensor at the bottom representing a suitable state~$\ket{\psi}\in \mathbb{C}^d$  fully determine the state under consideration. 
 \label{fig:homogeneousbinarytree}}
\end{figure}

In the general language of planar algebras, Jones~\cite{vaughanjonesunitaryreps} uses this type of maps (associated, in his context, with a special ``isometric'' element $V$ of the planar algebra) to define a Hilbert space~$\cH$ he calls the {\em semicontinuous limit}.
To briefly sketch this construction (specialised to the tensor planar algebra), recall that an interval $[a,b]\subseteq [0,1]$ is called {\em standard dyadic} if $a=p/2^n$ and $b=(p+1)/2^n$ for integers $p$ and $n$. 
Let $\cI$ be a partition of $[0,1]$ (which we think of as $S^1$) into standard dyadic intervals. Such a partition determines a rooted bifurcating tree~$\cT_{\cI}$ in $[0,1]\times [0,1]$ 
with root at $(0,1/2)$ and leaves at $\cM_\cI\times \{1\}$, were $\cM_\cI$ are the set of midpoints of intervals of $\cI$. The points~$\cM_\cI$ can be thought of as locations of spin-degrees of freedom on~$S^1$. Let us denote by $\cH_\cI:=(\mathbb{C}^d)^{\otimes |\cM_\cI|}$ the associated Hilbert space. The tree~$\cT_{\cI}$ together with an isometry~$V:\mathbb{C}^d\rightarrow\mathbb{C}^d\otimes \mathbb{C}^d$ determine an isometry~$\iota_{\cI}^{V}:\mathbb{C}^d\rightarrow\cH_\cI$. Thus any pair $(\cI,\ket{\psi})$, where $\ket{\psi}\in \mathbb{C}^d$ is a state, determines a state $\iota_{\cI}^{V}\ket{\psi}\in\cH_\cI$ via the associated tensor network, together with the ``locations'' $\cM_\cI$ of the spin degrees of freedom. 
The set~$\cD$ of standard dyadic partitions forms a directed set by inclusion: for $\cI,\cJ\in\cD$, we write $\cI\preceq\cJ$ if every interval of $\cI$ is the union of some intervals of $\cJ$. In this case, $\cJ$ is a ``refinement'' of $\cI$, and there is a natural associated (isometric) refining map $\iota_{\cJ,\cI}^V:\cM_\cI\rightarrow\cM_\cJ$. The map~$\iota_{\cJ,\cI}^V$ is (similarly as $\iota_{\cI}^{V}$) defined by a ``forest'', a certain graph with associated tensor network. In particular, this means that two
states $\Psi\in\cH_\cI$  and $\Phi\in\cH_\cJ$ can be compared by bringing them to  some (potentially finer) ``scale'' $\cK$, where $\cI\preceq\cK$ and $\cJ\preceq \cK$. One then considers the disjoint union of the spaces $\{\cH_{\cI}\}_{\cI\in\cD}$ and  imposes the equivalence relation $\Psi\sim \Phi$ (for $\Psi\in\cH_\cI$ and $\Phi\in\cH_\cJ$) if and only if  $\iota_{\cK,\cI}\Psi=\iota_{\cK,\cJ}\Phi$ for some~$\cK$ refining both $\cI$ and $\cJ$. The equivalence class of a state~$\Psi$ can thus be considered as the UV completion of~$\Psi$ (see~\cite{DynamicsHoloCodes}, where this construction is used in the context of holographic codes). Finally, inner products of such equivalence classes are taken by fine-graining representatives until they reach they same scale.
The semicontinuous limit Hilbert space~$\cH$ is then obtained by taking formal linear combinations of equivalence classes and the completion with respect to the induced norm.

Remarkably, as shown by Jones~\cite{vaughanjonesunitaryreps}, one can define a unitary action~$\rho^V$ of the Thompson group~$\thompson$ on the space~$\cH$ for any isometry $V:\mathbb{C}^d\rightarrow\mathbb{C}^d\otimes\mathbb{C}^d$. An element~$f$ of~$\thompson$ is a piecewise linear homeomorphism from $[0, 1]$ (with $0$ and $1$ identified such that $[0,1]\cong S^1$) to itself which is everywhere differentiable except at finitely many dyadic rational numbers and with slopes that are integer powers of~2 wherever~$f$ is differentiable.
The result of applying $\rho^V(f)$ to $[\Psi]$, where  $\Psi\in\cH_I$  can easily be described by a tensor network if the image $\cJ:=f(\cI)$ itself a standard dyadic partition: in this case, $\Psi$ and $f$ give rise to an element~$\Psi'\in\cH_{\cJ}$ simply by relabeling tensor product factors according to $f$, and~$\rho^V(f)[\Psi]:=[\Psi']$ is defined as the equivalence class of that element. If $f(\cI)$ is not a standard partition, one first has to refine: it can be shown~\cite{CFP,vaughanjonesunitaryreps} that for any $\cI\in\cD$ and $f\in\thompson$, there is some $\cI'\in\cD$ such that $\cI\preceq\cI'$ and $f(\cI')\in\cD$. Then $\rho^V(f)[\Psi]$ can be obtained by computing $\rho^V(f)[\iota_{\cI',\cI}\Psi]$. The fact that this is well-defined is non-trivial and involves a certain parametrization of the elements of~$\thompson$ as pairs of binary rooted trees~\cite{CFP}.

The interest in representations of~$\thompson$ stems from the fact that $\thompson$ is dense in the space~$\mathsf{Diff}^+(S^1)$ of orientation-preserving diffeomorphisms of $S^1$ (see~\cite[Proposition 5.2.]{DynamicsHoloCodes}~for a proof sketch). As a consequence, if $\rho^V$ were continuous in a suitable sense, this would naturally yield a representation of $\mathsf{Diff}^+(S^1)$, hence completing the translation from discrete (spin) degrees of freedom to the language of CFTs.

Here we show that  -- unfortunately -- this is generically not the case: in fact, the resulting representation~$\rho^V$ is not even weakly continuous for a typical choice of isometry~$V$.  The representation $\rho^V$ is called {\em weakly discontinuous at the identity} if there is a sequence~$\{f_k\}_{k\in\mathbb{N}}\subset \thompson$ with
\begin{align}
\lim_{k\rightarrow\infty}\|f_k-\mathsf{id}\|_\infty&=0\qquad\textrm{and }\label{eq:convergenceidentity}\\
\lim_{k\rightarrow\infty } \langle\Phi, \rho^V(f_k)\Psi\rangle&\neq \langle\Phi,\Psi\rangle \qquad\textrm{ for some }\ket{\Phi},\ket{\Psi}\in\cH\ .
\end{align}
Our main result is the  following no-go result: 
\begin{theorem}\label{thm:main}
For all but a zero-measure set of isometries~$V:\mathbb{C}^d\rightarrow\mathbb{C}^d\otimes\mathbb{C}^d$, the representation~$\rho^V$ is weakly discontinuous at the identity.
\end{theorem}
This kind of no-go theorem was first established by Jones~\cite{JonesNoGo} in the special case where $V$ is a certain   (in that case unique) isometric element of a specific planar algebra derived from the Temperley-Lieb algebra. The planar algebra can alternatively be interpreted as that  for the $3$-dimensional representation of the quantum group $U_qSO(3)$ (see~\cite{Jonesscaleinvarianttransfer}).  In contrast, our analysis is for arbitrary, and in particular, generic isometries in the tensor planar algebra, and is thus closer to tree tensor network descriptions of critical quantum spin chains.

Let us now present the proof of our main result. Given an isometry $V:\mathbb{C}^d\rightarrow \mathbb{C}^d\otimes\mathbb{C}^d$, we consider the projection $P:=VV^\dagger$ and its image $\cV=P(\mathbb{C}^d\otimes\mathbb{C}^d)$. We show the following
property:
\begin{lemma}\label{lem:necessarycondition}
If $\cV$ satisfies 
\begin{align}
(\cV\otimes\mathbb{C}^d)\cap (\mathbb{C}^d\otimes \cV)=\{0\}\ ,\label{eq:intersectioncondition}
\end{align}
then $\rho^V$ is weakly discontinuous at the identity.
\end{lemma}

Lemma~\ref{lem:necessarycondition} immediately implies Theorem~\ref{thm:main}
using algebraic geometry, which have been used in a similar fashion in the context of quantum satisfiability problems~\cite{laumannetalqsat2010}).   Indeed, suppose $V$ is defined by the first $d$ columns of a Haar-random unitary matrix~$U\in \mathsf{U}(d^2)$. Then the projection $P=VV^\dagger$ takes the form $P=P(U)=UP_0U^\dagger$ for a fixed rank-$d$ projection $P_0$ on $\mathbb{C}^d\otimes \mathbb{C}^d$. Let $\cW\subset \mathsf{U}(d^2)$ be the subset of unitaries 
such that the associated subspace $\cV:=P(\mathbb{C}^d\otimes\mathbb{C}^d)$ satisfies~\eqref{eq:intersectioncondition}. We will show that $\cW$ has full measure in the unitary group~$\mathsf{U}(d^2)$.

First observe that the operator
\begin{align}
\Gamma_P:=(P\otimes I_{\mathbb{C}^d})(I_{\mathbb{C}^d}\otimes P)(P\otimes I_{\mathbb{C}^d})
\end{align}
is Hermitian and satisfies $ 0\leq \Gamma_P \leq I_{(\mathbb{C}^d)^{\otimes 3}}$, where the latter inequality is strict if and only if~\eqref{eq:intersectioncondition} holds (or equivalently if $U\in\cW$).  In particular,
the operator $I_{(\mathbb{C}^d)^{\otimes 3}}-\Gamma_P$ is invertible if and only if $U\in\cW$. This means that the complement $\cW^c:=\mathsf{U}(d^2)\backslash \cW$ is the set of zeros of the function $f(U):=\det(I_{(\mathbb{C}^d)^{\otimes 3}}-\Gamma_{P(U)})$. This function is a polynomial in the real- and imaginary parts of the matrix entries of~$U$. 
By using algebraic geometry (more explicitly, we use~\cite[Lemma 4.3]{NechitaPellegrini12}), we conclude that~$\cW^c$ is either the whole set $\mathsf{U}(d^2)$ or a set of zero measure.  It remains to argue that $\cW^c\neq \mathsf{U}(d^2)$: for this, we explicitly demonstrate the existence of a unitary~$U\in\cW$. Clearly, it suffices to show existence of a $d$-dimensional subspace~$\cV$ of $\mathbb{C}^d\otimes\mathbb{C}^d$ satisfying~\eqref{eq:intersectioncondition}.  Define a stabilizer operator $S=X\otimes Z$, where $X$ and $Z$ are the generalized Pauli operators and let $\cV$ be the subspace stabilized by~$S$. Then $\dim(\cV)=d$, and 
$(\cV\otimes \mathbb{C}^d)\cap (\mathbb{C}^d\otimes \cV)$ is stabilized by $S_1=S\otimes I_{\mathbb{C}^d}$ and $S_2=I_{\mathbb{C}^d}\otimes S$. Since $S_1$ and $S_2$ do not commute, this implies~\eqref{eq:intersectioncondition}.

It remains to show Lemma~\ref{lem:necessarycondition}. Assume that an isometry~$V$ is given such that the associated space~$\cV$ satisfies~\eqref{eq:intersectioncondition}. Similar to the argument in Jones~\cite{JonesNoGo}, we use the sequence~$\{f_k\}_{k\in \mathbb{N}}$ of translations by~$\frac{1}{2^k}$ to show that $\rho^V$ is weakly discontinuous at the identity. Clearly, this sequence converges to the identity in the sense of~\eqref{eq:convergenceidentity}. Consider two arbitrary non-orthogonal states $\psi,\phi\in\mathbb{C}^d$. These define states~$\ket{\Phi},\ket{\Psi}\in\cH$ as described above with $\langle\Phi,\Psi\rangle\neq 0$. We will show that $\lim_{k\rightarrow \infty }\langle \Phi ,\rho^V(f_k)\Psi\rangle=0$.

\begin{figure}
\centering
    \subfigure[\label{fig:topleft} ]{
   
\begin{tikzpicture}[scale = 0.2]
\draw [semithick] (0, 5.5) -- (0,-1.5) -- (0.5,-2) -- (1, -1.5) -- (1, -1) -- (1.5, -0.5) -- (2, -1) -- (2, -4.5) -- (2.5, -5) -- (3, -4.5) -- (3, -4) -- (3.5, -3.5) 
-- (4, -4) -- (4, -11.5) -- (4.5, -12) -- (5, -11.5) -- (5, -11) -- (5.5, -10.5) -- (6, -11) -- (6, -14.5) -- (6.5, -15) -- (7, -14.5) -- (7, -14) -- (7.5, -13.5) 
-- (8, -14) -- (8, -21);

\draw [fill=darkgreen, semithick] (0, -1.5) -- (0.5, -2) -- (1, -1.5) -- (0, -1.5);
\draw [fill=darkgreen, semithick] (1, -1) -- (2, -1) -- (1.5, -0.5) -- (1, -1);
\draw [fill=darkgreen, semithick] (2, -4.5) -- (2.5, -5) -- (3, -4.5) -- (2, -4.5);
\draw [fill=darkgreen, semithick] (3, -4) -- (4, -4) -- (3.5, -3.5) -- (3, -4);
\draw [fill=darkgreen, semithick] (4, -11.5) -- (4.5, -12) -- (5, -11.5) -- (4, -11.5);
\draw [fill=darkgreen, semithick] (5, -11) -- (6, -11) -- (5.5, -10.5) -- (5, -11);
\draw [fill=darkgreen, semithick] (6, -14.5) -- (6.5, -15) -- (7, -14.5) -- (6, -14.5);
\draw [fill=darkgreen, semithick] (7, -14) -- (8, -14) -- (7.5, -13.5) -- (7, -14);

\draw [semithick] (0.5, -2) -- (0.5, -6) -- (1.5, -7) -- (2.5, -6) -- (2.5, -5); 
\draw [semithick] (4.5, -12) -- (4.5, -16) -- (5.5, -17) -- (6.5, -16) -- (6.5, -15); 

\draw [fill=darkgreen, semithick] (1, -6.5) -- (1.5, -7) -- (2, -6.5) -- (1, -6.5);
\draw [fill=darkgreen, semithick] (5, -16.5) -- (5.5, -17) -- (6, -16.5) -- (5, -16.5);

\draw [semithick] (1.5, -0.5) -- (1.5, 0.5) -- (2.5, 1.5) -- (3.5, 0.5) -- (3.5, -3.5); 
\draw [semithick] (5.5, -10.5) -- (5.5, -9.5) -- (6.5, -8.5) -- (7.5, -9.5) -- (7.5, -13.5); 

\draw [fill=darkgreen, semithick] (2, 1) -- (2.5, 1.5) -- (3, 1) -- (2, 1);
\draw [fill=darkgreen, semithick] (4, 4) -- (4.5, 4.5) -- (5, 4) -- (4, 4);

\draw [semithick] (1.5, -7) -- (1.5, -18) -- (3.5, -20) -- (5.5, -18) -- (5.5, -17); 
\draw [semithick] (2.5, 1.5) -- (2.5, 2.5) -- (4.5, 4.5) -- (6.5, 2.5) -- (6.5, -8.5); 

\draw [fill=darkgreen, semithick] (3, -19.5) -- (3.5, -20) -- (4, -19.5) -- (3, -19.5);
\draw [fill=darkgreen, semithick] (6, -9) -- (6.5, -8.5) -- (7, -9) -- (6, -9);

\draw [thick, dotted] (-0.5, 0) rectangle (2.5, -2.5);  
\draw [thick, dotted] (1.5, -3) rectangle (4.5, -5.5);  
\draw [thick, dotted] (3.5, -10) rectangle (6.5, -12.5);  
\draw [thick, dotted] (5.5, -13) rectangle (8.5, -15.5);  

\draw [semithick] (4.5, 4.5) -- (4.5, 5.5); 
\draw [semithick] (3.5, -21) -- (3.5, -20); 

\draw [semithick] (8, -21) arc (0:-90:3cm);
\draw [semithick] (0, -24) -- (5, -24); 
\draw [semithick] (0, -24) arc (-90:-180:3cm);
\draw [semithick] (-3, -21) -- (-3, 5.5); 
\draw [semithick] (-3, 5.5) arc (180:0:1.5cm);

\node[above] at (4.5, 5.1) {$\psi$};
\node[below] at (3.5, -20.6) {$\phi$};
\end{tikzpicture} 
    }\qquad
    \subfigure[\label{fig:topright}]{
    \begin{tikzpicture}[scale = 0.2]
    

\draw [semithick] (17, 5.5) -- (17, 0); 
\draw [semithick] (25, -15.5) -- (25, -21); 

\draw [semithick] (18.5, 0) -- (18.5, 0.5) -- (19.5, 1.5) -- (20.5, 0.5) -- (20.5, -3); 
\draw [semithick] (22.5, -10) -- (22.5, -9.5) -- (23.5, -8.5) -- (24.5, -9.5) -- (24.5, -13); 
\draw [semithick] (17.5, -2.5) -- (17.5, -6) -- (18.5, -7) -- (19.5, -6) -- (19.5, -5.5); 
\draw [semithick] (21.5, -12.5) -- (21.5, -16) -- (22.5, -17) -- (23.5, -16) -- (23.5, -15.5); 

\draw [semithick] (18.5, -7) -- (18.5, -18) -- (20.5, -20) -- (22.5, -18) -- (22.5, -17); 
\draw [semithick] (19.5, 1.5) -- (19.5, 2.5) -- (21.5, 4.5) -- (23.5, 2.5) -- (23.5, -8.5); 

\draw [thick] (16.5, 0) rectangle (19.5, -2.5);  
\draw [thick] (18.5, -3) rectangle (21.5, -5.5);  
\draw [thick] (20.5, -10) rectangle (23.5, -12.5);  
\draw [thick] (22.5, -13) rectangle (25.5, -15.5);  
\node at (18, -1.25) {$x$};
\node at (20, -4.25) {$x$};
\node at (22, -11.25) {$x$};
\node at (24, -14.25) {$x$};

\draw [semithick] (21.5, 4.5) -- (21.5, 5.5); 
\draw [semithick] (20.5, -21) -- (20.5, -20); 
\node[above] at (21.5, 5.1) {$\psi$};
\node[below] at (20.5, -20.6) {$\phi$};
\draw [semithick] (19, -2.5) -- (19, -3); 
\draw [semithick] (21, -5.5) -- (21, -10); 
\draw [semithick] (23, -12.5) -- (23, -13); 

\draw [thick, dotted] (16, 2) rectangle (22, -7.5);  
\draw [thick, dotted] (20, -8) rectangle (26, -17.5);  

\draw [fill=darkgreen, semithick] (18, -6.5) -- (18.5, -7) -- (19, -6.5) -- (18, -6.5);
\draw [fill=darkgreen, semithick] (22, -16.5) -- (22.5, -17) -- (23, -16.5) -- (22, -16.5);

\draw [fill=darkgreen, semithick] (19, 1) -- (19.5, 1.5) -- (20, 1) -- (19, 1);
\draw [fill=darkgreen, semithick] (21, 4) -- (21.5, 4.5) -- (22, 4) -- (21, 4);

\draw [fill=darkgreen, semithick] (20, -19.5) -- (20.5, -20) -- (21, -19.5) -- (20, -19.5);
\draw [fill=darkgreen, semithick] (23, -9) -- (23.5, -8.5) -- (24, -9) -- (23, -9);

\draw [semithick] (25, -21) arc (0:-90:3cm);
\draw [semithick] (17, -24) -- (22, -24); 
\draw [semithick] (17, -24) arc (-90:-180:3cm);
\draw [semithick] (14, -21) -- (14, 5.5); 
\draw [semithick] (14, 5.5) arc (180:0:1.5cm);
    \end{tikzpicture}
    }\\
     \subfigure[\label{fig:bottomleft}]{
         \begin{tikzpicture}[scale = 0.2]
\draw [semithick] (0, -26.5) -- (0, -30); 
\draw [semithick] (8, -49.5) -- (8, -53); 
\draw [semithick] (4, -39.5) -- (4, -40); 
\draw [semithick] (4.5, -27.5) -- (4.5, -26.5); 
\draw [semithick] (3.5, -52) -- (3.5, -53); 

\draw [semithick] (2.5, -30) -- (2.5, -29.5) -- (4.5, -27.5) -- (6.5, -29.5) -- (6.5, -40); 
\draw [semithick] (1.5, -39.5) -- (1.5, -50) -- (3.5, -52) -- (5.5, -50) -- (5.5, -49.5); 

\draw [thick] (-1, -30) rectangle (5, -39.5);  
\draw [thick] (3, -40) rectangle (9, -49.5);  
\node at (2, -34.75) {$\cR(x)$};
\node at (6, -44.75) {$\cR(x)$};

\draw [thick, dotted] (-1.5, -27) rectangle (9.5, -52.5);  
\node[above] at (4.5, -26.9) {$\psi$};
\node[below] at (3.5, -52.6) {$\phi$};

\draw [fill=darkgreen, semithick] (4, -28) -- (4.5, -27.5) -- (5, -28) -- (4, -28);
\draw [fill=darkgreen, semithick] (3, -51.5) -- (3.5, -52) -- (4, -51.5) -- (3, -51.5);

\draw [semithick] (8, -53) arc (0:-90:3cm);
\draw [semithick] (0, -56) -- (5, -56); 
\draw [semithick] (0, -56) arc (-90:-180:3cm);
\draw [semithick] (-3, -53) -- (-3, -26.5); 
\draw [semithick] (-3, -26.5) arc (180:0:1.5cm);
        \end{tikzpicture}
    }\qquad
    \subfigure[ \label{fig:bottomright}]{
    \begin{tikzpicture}[scale=0.2]
    
\draw [thick] (15.5, -27) rectangle (26.5, -52.5);  

\draw [semithick] (25, -52.5) -- (25, -53); 
\draw [semithick] (25, -53) arc (0:-90:3cm);
\draw [semithick] (17, -56) -- (22, -56); 
\draw [semithick] (17, -56) arc (-90:-180:3cm);
\draw [semithick] (14, -53) -- (14, -26.5); 
\draw [semithick] (14, -26.5) arc (180:0:1.5cm);
\draw [semithick] (17, -26.5) -- (17, -27); 

\draw [semithick] (21.5, -27) -- (21.5, -26.5); 
\draw [semithick] (20.5, -52.5) -- (20.5, -53); 
\node[above] at (21.5, -26.9) {$\psi$};
\node[below] at (20.5, -52.6) {$\phi$};

\node at (21, -39.75) {$\cR^{\circ 2}(x)$};
        \end{tikzpicture}
    
    }
    
\caption[Inner Product]{
The inner product $\langle \Phi ,\rho^V(f_3)\Psi\rangle$. Here we assume periodic boundary conditions.  
After $3$ layers of application of
the fine-graining isometry~$V$, each state
is represented by an $8$-qudit state. The inner product is obtained
by translating the second state~$\Psi$ by~$1/2^3$ (which corresponds to one lattice spacing if the qudits are arranged equidistantly on the unit circle), then contracting the tensor network obtained by 
stacking the adjoing of $\Phi$ on top. In Fig.~\ref{fig:topleft},
appearances of the tensor~$x$
(see  Fig.~\ref{fig:x}) are indicated by dotted boxes. 
In Fig.~\ref{fig:topright}  the same inner product is shown with $x$ inserted,
and with dotted boxes indicating the position of $\cR(x)$ (see  Fig.~\ref{fig:rz}).
 Fig.~\ref{fig:bottomleft} shows the same inner product with $\cR(x)$ inserted; now, the dotted boxes indicate the position of $\cR^{\circ 2}(x)$. 
Finally,  Fig.~\ref{fig:bottomright} shows the final form of the inner product which corresponds to the right hand side of Eq. \eqref{eq:cRAexplicit}.}
\label{fig:ScalarProduct}
     \end{figure}
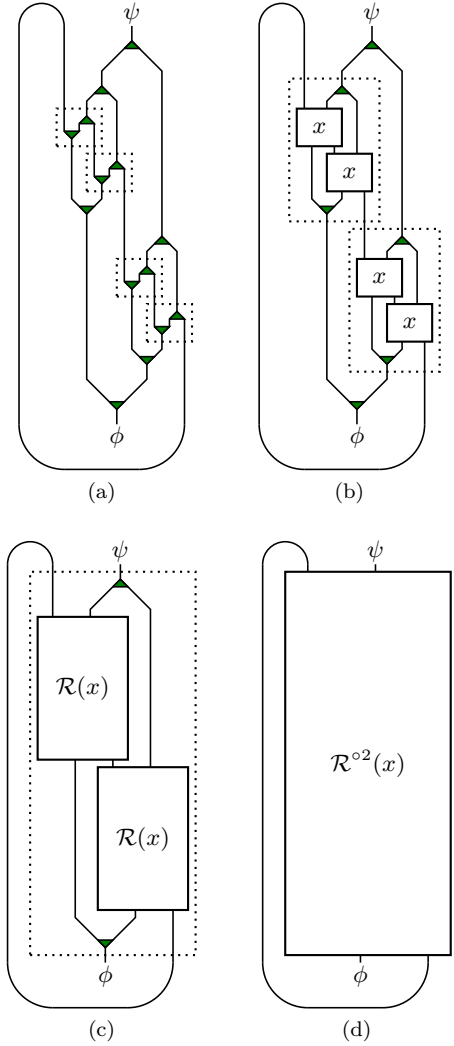

As shown in Fig.~\ref{fig:ScalarProduct}, the matrix element $\langle \Phi ,\rho^V(f_k)\Psi\rangle$ can be expressed as 
\begin{align}
\langle \Phi ,\rho^V(f_k)\Psi\rangle &=\tr\left(\cR^{\circ k-1}(x)A(\phi,\psi)\right)\ .\label{eq:cRAexplicit}
\end{align}

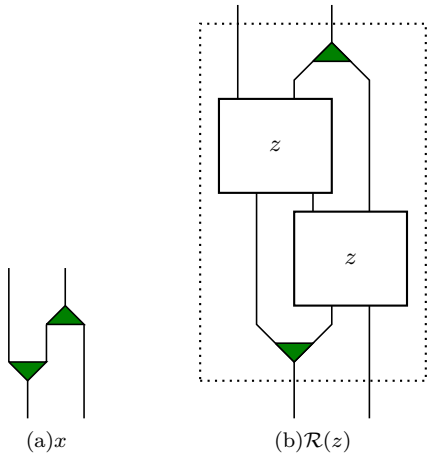
\begin{figure}
    \centering
    \subfigure[ $x$\label{fig:x}]
{ \begin{tikzpicture}[scale=0.5]
\draw [semithick] (0,4) -- (0,1.5); 
\draw [fill=darkgreen, semithick] (0,1.5) -- (0.5,1) -- (1,1.5) -- (0,1.5);
\draw [semithick] (1,1.5) -- (1,2.5); 
\draw [fill=darkgreen, semithick] (1,2.5) -- (2,2.5) -- (1.5,3) -- (1,2.5);
\draw [semithick] (2,2.5) -- (2,0); 
\draw [semithick] (0.5,1) -- (0.5,0);
\draw [semithick] (1.5,3) -- (1.5,4);

\end{tikzpicture}
}\qquad\qquad
 \subfigure[ $\mathcal{R}(z)$\label{fig:rz}]
{
\begin{tikzpicture}[scale=0.5]
\draw [semithick] (17,2.5) -- (17, 0); 
\draw [semithick] (18.5, 0) -- (18.5, 0.5) -- (19.5, 1.5) -- (20.5, 0.5) -- (20.5, -3); 
\draw [semithick] (17.5, -2.5) -- (17.5, -6) -- (18.5, -7) -- (19.5, -6)-- (19.5, -5.5); 
\draw [semithick] (19.0, -2.5)-- (19.0, -3);
\draw [semithick] (18.5, -7) -- (18.5, -8.5);
\draw [semithick] (19.5, 1.5) -- (19.5, 2.5);
\draw [thick] (16.5, 0) rectangle (19.5, -2.5);  
\draw [thick] (18.5, -3) rectangle (21.5, -5.5);  
\node at (18, -1.25) {$z$};
\node at (20, -4.25) {$z$};
\draw [semithick] (20.5, -5.5) -- (20.5, -8.5); 
\draw [thick, dotted] (16, 2) rectangle (22, -7.5);  
\draw [fill=darkgreen, semithick] (18, -6.5) -- (18.5, -7) -- (19, -6.5) -- (18, -6.5);
\draw [fill=darkgreen, semithick] (19, 1) -- (19.5, 1.5) -- (20, 1) -- (19, 1);
\end{tikzpicture}
}
\caption{The element $x\in\cB(\mathbb{C}^d\otimes\mathbb{C}^d)$ (cf.~\eqref{eq:xdefinition}) and the definition of $\cR(z)\in\cB(\mathbb{C}^d\otimes\mathbb{C}^d)$ (cf.~\eqref{eq:rzdefinition})  used in the proof of Lemma~\ref{lem:necessarycondition}.}
\end{figure}

     \noindent Here $A(\phi,\psi)\in \cB(\mathbb{C}^d\otimes\mathbb{C}^d)$ is an operator independent of~$k$, the operator 
\begin{align}
x:= (I_{\mathbb{C}^d}\otimes V^\dagger) (V\otimes I_{\mathbb{C}^d})\in \cB(\mathbb{C}^d\otimes\mathbb{C}^d)\label{eq:xdefinition}
\end{align} is defined in terms of the isometry~$V$ (see Figure \ref{fig:x}), and
$\cR:\cB(\mathbb{C}^d\otimes\mathbb{C}^d)\rightarrow\cB(\mathbb{C}^d\otimes\mathbb{C}^d)$ is a superoperator quadratic in the argument~$z$ given by  (see Fig.~\ref{fig:rz})
\begin{align}
\cR(z)&=(I_{\mathbb{C}^d}\otimes V^\dagger)(z\otimes I_{\mathbb{C}^d})(I_{\mathbb{C}^d}\otimes z) (V\otimes I_{\mathbb{C}^d})\ .\label{eq:rzdefinition}
\end{align}

\noindent We will show that the operator norm  of~$\cR^{\circ k}(x)$ decays doubly exponentially with~$k$ such that 
\begin{align}
\lim_{k\rightarrow\infty } \|\cR^{\circ k}(x)\|=0\ .
\end{align}
The claim then follows from~\eqref{eq:cRAexplicit}. 

More precisely, we claim that 
\begin{align}
\|\cR^{\circ k}(x)\|\leq \|(I\otimes V^\dagger)(V\otimes I)\|^{ 2^{k}}\quad\textrm{ for all }k\in\mathbb{N}_0\ \label{eq:doubleexponentialdecay}
\end{align}
where we write $I$ for the identity on~$\mathbb{C}^d$ for ease of notation. 
Since $V\otimes I=(P\otimes I)(V\otimes I)$ by definition of $P$ and the operator norm is unchanged when applying the isometry $I\otimes V$ on the left, we obtain 
\begin{align}
\|(I\otimes V^\dagger)(V\otimes I)\|
&= \|(I\otimes V)(I\otimes V^\dagger)(P\otimes I)(V\otimes I)\|\nonumber\\
&=\|(I\otimes P)(P\otimes I)(V\otimes I)\|\nonumber\\
&\leq\|(I\otimes P)(P\otimes I)\| <1\ .\label{eq:upperboundv}
\end{align}
Here we used the submultiplicativity property $\|AB\|\leq \|A\|\cdot\|B\|$
of the operator norm and the fact that 
$|W\|=\|W^\dagger\|=1$ for any isometry~$W$ to obtain the inequality, and our assumption~\eqref{eq:intersectioncondition} in the last step.

 To show~\eqref{eq:doubleexponentialdecay}, observe that for $k=0$, we have 
\begin{align}
\|R^{\circ 0}(x)\|=\|x\|&=\|(I\otimes V^\dagger)(V\otimes I)\|<1
\end{align}
by~\eqref{eq:upperboundv}. For $k\geq 0$ we obtain
\begin{align}
\|\cR^{\circ k+1}(x)\|&=\|\cR\circ \cR^{\circ k}(x)\|\\
&=\|(I \otimes V^\dagger)(\cR^{\circ k}(x) \otimes I )(I \otimes \cR^{\circ k}(x)) (V\otimes I)\|\\
&\leq \|(I\otimes V)^\dagger\|\cdot \|\cR^{\circ k}(x)\|^2\cdot \|I\otimes V\|\ \\
&=\|\cR^{\circ k}(x)\|^2
\end{align}
by the submultiplicativity and the stability property $\|A\otimes I\|=\|A\|$ of the operator norm. This proves \eqref{eq:doubleexponentialdecay} and therefore Lemma \ref{lem:necessarycondition}.

As a  non-trivial example where the condition~\eqref{eq:intersectioncondition} of Lemma~\ref{lem:necessarycondition}
does not apply, consider the case of spin-chains invariant under a global symmetry group~$G$, as described by a unitary representation~$g\mapsto U_g\in\mathsf{U}(d)$. In this case, the isometry $V$ has the symmetry property (see e.g.,~\cite{Huangetal13,SinghVidal13})
\begin{align}
V=(U_g\otimes U_g)VU_g^\dagger\qquad\textrm{for all }g\in G\ .\label{eq:isometryv}
\end{align}
Concretely, in the case $d=3$, we may consider the spin-$1$ irreducible representation of $SO(3)$ with angular momentum eigenbasis $\ket{\text{-}1},\ket{0},\ket{1}$.
Since the tensor product of the two spin-$1$ representations decomposes as~$1\otimes 1=0\oplus 1\oplus 2$, the symmetry~\eqref{eq:isometryv} and dimensionality considerations show that $\cV$ is the subspace associated with the spin-$1$ representation, and the isometry~$V$ is unique up to a global phase (as in the example of Jones). The space is spanned by the vectors
\begin{align}
|J_z=1\rangle &=\frac{1}{\sqrt{2}} (\ket{1,0}-\ket{0,1})\\
|J_z=0\rangle &=\frac{1}{\sqrt{2}} (\ket{1,\text{-}1}-\ket{\text{-}1,1})\\
|J_z=-1\rangle &=\frac{1}{\sqrt{2}} (\ket{0,\text{-}1}-\ket{\text{-}1,0})\ .
\end{align}
From this it is easy to check that the space~$(\cV\otimes\mathbb{C}^d)\cap (\mathbb{C}^d\otimes \cV)$
is $1$-dimensional and spanned by
the vector $-\ket{\text{-}1,0,1}+\ket{\text{-}1,1,0}+\ket{0,\text{-}1,1}-\ket{0,1,\text{-}1}-\ket{1,\text{-}1,0}+\ket{1,0,\text{-}1}$. Hence we conclude that Lemma~\ref{lem:necessarycondition} does not apply in this case. It remains an open problem to establish continuity or discontinuity of the associated representation.

\emph{Conclusions.}
We have shown that Jones' approach to constructing a chiral conformal field theory 
whose Hamiltonian generates translations on a circle as a limit of rotations on a discrete lattice fails
for a generic choice of isometry~$V$ as this limit is discontinuous. Our result leaves open the possibility that there exists a continuous representation $\rho^V$ for an appropriately chosen isometry~$V$, but  provides a simple necessary condition for continuity (see Lemma~\ref{lem:necessarycondition}). However, it shows that even if such an isometry exists, the corresponding renormalization group fixed point  would be unstable: small perturbations in~$V$ would suffice to produce the discontinuity shown above. Thus our result suggests that tree tensor networks are insufficient to capture dynamical aspects of CFTs, at least following the approach by Jones. 

Given this no-go-statement, it is natural to look for other ways of defining a semicontinuous limit of Hilbert spaces: one way proposed by Jones is to consider additional local ``disentanglers'' between spins in the spirit of the MERA~\cite{PhysRevLett.99.220405}. An alternative would be to consider anyonic spin chains derived from modular tensor categories (see also~\cite{jonesqgroups}), where the resulting state space can have fractional dimension.
Candidate isometries having the ``perfectness'' property suitable for use in the AdS/CFT-correspondence 
have recently been identified~\cite{bergerosborne}. However, establishing continuity of such representations (even for the above example based on $SO(3)$) remains a challenge.

\emph{Acknowledgments.}
 RK  is supported by the Technische Universit\"at M\"unchen -- Institute  for  Advanced  Study,  funded  by  the  German  Excellence  Initiative  and  the European  Union  Seventh  Framework  Programme  under  grant  agreement  no.~291763. He thanks Ion Nechita, David Gosset and Sergey Bravyi
 for helpful discussions concerning the algebraic geometry arguments, and Tobias Osborne for introducing him to Jones' work, as well as comments on the manuscript.
%


\end{document}